\documentclass[journal]{IEEEtran}
\usepackage{amsmath,amsfonts}
\usepackage{algorithmic}
\usepackage{algorithm}
\usepackage{array}
\usepackage{textcomp}
\usepackage{tabularx}
\usepackage{stfloats}
\usepackage{setspace} 
\usepackage{url}
\usepackage{verbatim}
\usepackage{graphicx}
\usepackage{cite}
\usepackage{multirow}
\usepackage{graphicx}
\usepackage{makecell}
\usepackage{booktabs}
\usepackage{tabularx}
\usepackage{CJK}
\usepackage{diagbox}
\usepackage{color}
\usepackage{colortbl} 
\usepackage{xcolor}
\usepackage{subfig}
\begin{document}
\begin{CJK}{UTF8}{gbsn}

\title{Audio Deepfake Attribution: An Initial Dataset and Investigation}

\author{Xinrui Yan,~\IEEEmembership{Student Member,~IEEE,}
        Jiangyan Yi,~\IEEEmembership{Member,~IEEE,}
        Jianhua Tao,~\IEEEmembership{Senior Member,~IEEE,}  
        \newline
        Jie Chen,~\IEEEmembership{Member,~IEEE}
\thanks{Xinrui Yan and Jie Chen are with the School of Electronics and Information Engineering, Beihang University, Beijing, China. (e-mail: yanxinrui2021@ia.ac.cn, chenjie@buaa.edu.cn)

Jiangyan Yi is with the Department of Automation, Tsinghua University, Beijing, China. (e-mail: jiangyan.yi@nlpr.ia.ac.cn)

Jianhua Tao is with the Department of Automation, Tsinghua University, Beijing, China and National Research Center for Information Science and Technology, Tsinghua University, Beijing, China. (e-mail: jhtao@tsinghua.edu.cn)

Corresponding Author: Jiangyan Yi, Jianhua Tao. (e-mail: jiangyan.yi@nlpr.ia.ac.cn; 
jhtao@tsinghua.edu.cn)}}


\markboth{Journal of \LaTeX\ Class Files,~Vol.~14, No.~8, August~2021}%
{Shell \MakeLowercase{\textit{et al.}}: A Sample Article Using IEEEtran.cls for IEEE Journals}


\maketitle

\begin{abstract}
The rapid progress of deep speech synthesis models has posed significant threats to society such as malicious manipulation of content. This has led to an increase in studies aimed at detecting so-called ``deepfake audio". However, existing works focus on the binary detection of real audio and fake audio. In real-world scenarios such as model copyright protection and digital evidence forensics, binary classification alone is insufficient. It is essential to identify the source of deepfake audio. Therefore, audio deepfake attribution has emerged as a new challenge. To this end, we designed the first deepfake audio dataset for the attribution of audio generation tools, called Audio Deepfake Attribution (ADA), and conducted a comprehensive investigation on system fingerprints. To address the challenges of attribution of continuously emerging unknown audio generation tools in the real world, we propose the Class-Representation Multi-Center Learning (CRML) method for open-set audio deepfake attribution (OSADA). CRML enhances the global directional variation of representations, ensuring the learning of discriminative representations with strong intra-class similarity and inter-class discrepancy among known classes. Finally, the strong class discrimination capability learned from known classes is extended to both known and unknown classes. Experimental results demonstrate that the CRML method effectively addresses open-set risks in real-world scenarios.
The dataset is publicly available at:
https://zenodo.org/records/13318702, and https://zenodo.org/records/13340666.
\end{abstract}


\section{Introduction}
\IEEEPARstart{T}{ext} -to-speech (TTS), or speech synthesis aims to synthesize intelligible and natural speech given text \cite{ref1}. The rapid development of deep learning \cite{ref2} \cite{ref3} in recent years has made the technology of speech synthesis \cite{ref4} \cite{ref5} \cite{ref6} \cite{ref7} \cite{ref8} \cite{ref9} \cite{ref10} \cite{ref11} more and more mature. And it has been used in many commercial synthesis tools, such as Baidu Ai Cloud TTS system\footnote{https://ai.baidu.com/tech/speech/tts}, etc.

However, the malicious use and spread of this technology can be extremely destructive. With the popularity and speed of social media coverage, deepfake information can quickly reach many internet users. The hazards that exist include falsifying the audio of politicians to disrupt politics, falsifying court audio evidence, and telecommunication fraud. Therefore, verifying the authenticity of audio is crucial.

\begin{figure}[!t]
\centering
\includegraphics[width=\linewidth,height=4.5cm,keepaspectratio]{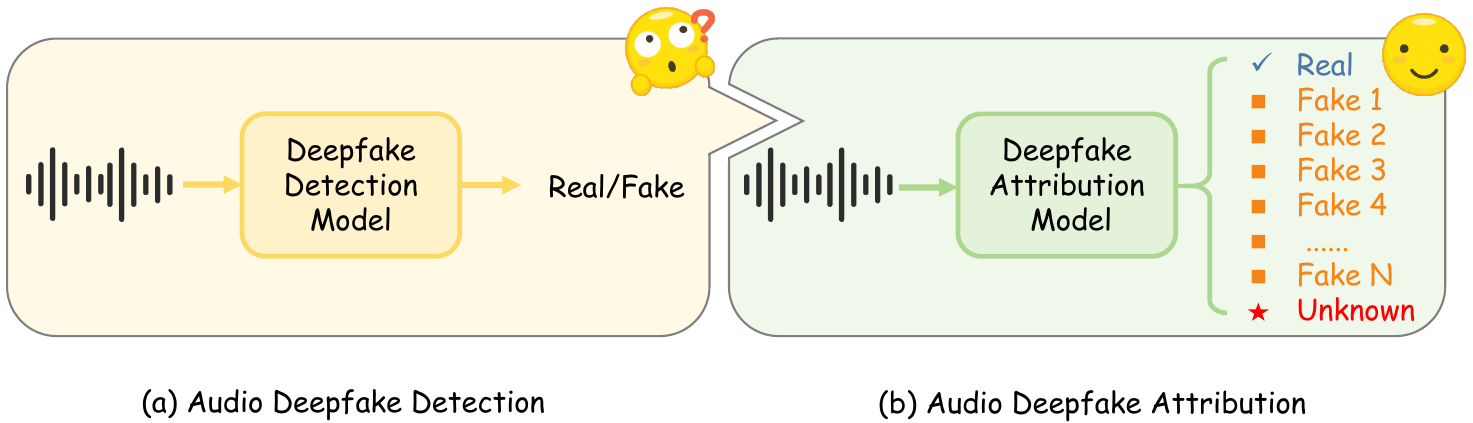}
\caption{Beyond the Limitation of Binary Deepfake Detection. (a) Previous audio deepfake detection focuses on distinguishing between real and fake audio. (b) Audio deepfake attribution aims to attribute audio generated by different deepfake technologies to their sources. Fake 1-Fake N refers to different audio generation algorithms or tools.}
\label{fig_2}
\end{figure}

\begin{table*}[ht]
\renewcommand\arraystretch{1.5} 
\centering
\caption{Information statistics of the existing deepfake audio datasets. Previous datasets have focused on the audio deepfake detection task, while our ADA dataset is designed for the audio deepfake attribution task. }
\large
\resizebox{\linewidth}{!}{
\begin{tabular}{cccccccc}
\toprule
Name & Year & Task & Language & Fake Types & Utterances & Speakers & Download URL \\  
\midrule
FoR \cite{ref25} & 2019 & Detection & English & TTS & 195,541 & \makecell{Fake:33\\Real:140} & \url{http://bil.eecs.yorku.ca/datasets} \\ 
WaveFake \cite{ref26} & 2021 & Detection & \makecell{English\\Japanese} & TTS & 117,985 & \makecell{Fake:2\\Real:2} & \url{https://zenodo.org/record/5642694} \\ 
ASVspoof 2021 DF \cite{ref15} & 2021 & Detection & English & \makecell{TTS, VC, \\Replay} & 1,566,273 & \makecell{Fake:133\\Real:133} & \url{https://doi.org/10.5281/zenodo.4835108} \\ 
FMFCC-A \cite{ref27} & 2021 & Detection & Chinese & \makecell{TTS, VC} & 50,000 & \makecell{Fake:73\\Real:58} & \url{https://github.com/Amforever/FMFCC-A} \\ 
ADD 2022 \cite{ref18} & 2022 & Detection & Chinese & \makecell{TTS, VC, \\Partially Fake} & 493,123 & Unknown & \url{http://addchallenge.cn/downloadADD2022} \\ 
\rowcolor{pink!20}
\textbf{ADA (Ours)} & \textbf{2022} & \textbf{Attribution} & \textbf{Chinese} & \textbf{TTS} & \textbf{181,764} & \textbf{\makecell{Fake:160\\Real:585}} & \textbf{\makecell[l]{https://zenodo.org/records/13318702\\https://zenodo.org/records/13340666}} \\ 
ADD 2023 \cite{ref19} & 2023 & \makecell{Detection \\ Localization \\ Attribution} & Chinese & \makecell{TTS, \\Partially Fake} & 517,068 & Unknown & \url{http://addchallenge.cn/downloadADD2023} \\ 
\bottomrule
\end{tabular}}
\end{table*}

To address this security issue, there has been a surge of interest and concern among researchers to develop audio deepfake detection methods \cite{ref12} \cite{ref13} \cite{ref14} \cite{ref15} \cite{ref16} \cite{ref17} \cite{ref18} \cite{ref19} \cite{ref20} \cite{ref62} \cite{ref63} \cite{ref64} \cite{ref65}. Among them, the availability of large-scale datasets is a favourable factor in the development of deepfake audio detection methods \cite{ref21} \cite{ref22} \cite{ref23} \cite{ref24}, as summarised in Table I. In 2019, Reimao et al. presented the FoR \cite{ref25} dataset for synthetic speech detection, which collects utterances from speech synthesis tools, including open-source and commercial methods. In 2021, Frank et al. presented the WaveFake \cite{ref26} dataset to facilitate audio deepfake detection, with samples from six different state-of-the-art architectures. ASVspoof2021 challenge \cite{ref16} added new speech deepfake (DF) task for deepfake speech detection. Zhang et al. presented a Mandarin dataset (FMFCC-A) \cite{ref27} for improving the detection of synthesized Mandarin speech. Considering many challenging attack situations in realistic scenarios, Yi et al. launched the Audio Deep synthesis Detection challenge (ADD 2022 and ADD 2023) \cite{ref18} \cite{ref19} to fill in the gap.

Previous research on audio deepfake detection has made significant contributions to the field. However, existing audio deepfake detection is primarily limited to binary classification, distinguishing between real and fake audio. 
This binary approach often lacks the capacity for further explanation and result analysis, which can undermine its convincingness. Solving this problem has essential significance for some realistic scenarios, mainly including (1) audio forensics \cite{ref28} \cite{ref29}: audio forensics includes different research, such as checking the authenticity of audio and providing judicial evidence in court; (2) Model property protection \cite{ref30} \cite{ref31}: a virtual model as a commercial service may be stolen. As valuable intellectual property, such models should be protected. In these scenarios, binary classification alone is insufficient. It is essential to identify the source of deepfake
audio. Consequently, audio deepfake attribution has emerged as a new challenge. Audio deepfake attribution aims to attribute audio generated by different deepfake technologies to their sources, including both the audio generation algorithms and the audio generation tools. This work focuses on audio generation tool attribution.

Advanced TTS tools from different vendors have provided tremendous practical benefits in current applications. These diverse tools are likely to produce unique ``system fingerprints", where the speech generated by each tool possesses certain distinctive and discernible characteristics. This prompts us to investigate system fingerprinting for deepfake audio, which attributes various deepfake audio samples to their respective generation tools.

To address this challenge, we designed the first deepfake audio dataset for
the attribution of audio generation tools, called Audio Deepfake Attribution (ADA), which includes audio synthesized by TTS systems from seven representative Chinese vendors. In the real world, deepfake audio inevitably undergoes data compression during online dissemination. To address this, we established a compressed dataset based on clean data, which includes unseen source and video format compression variants. Utilizing the ADA dataset, we are the first to conduct an investigation into audio deepfake attribution through system fingerprinting. 

As audio deepfake attribution models are deployed in real-world applications, new data categories, including deepfake audio from previously unknown generation tools, continually emerge. Therefore, the audio deepfake attribution system must reject unknown classes during testing, framing this as an open-set recognition (OSR) problem termed open-set audio deepfake attribution (OSADA). OSADA aims to recognize deepfake audio that does not belong to any known generation sources. To address the challenge of secure model deployment, we propose an open-set recognition method based on class-representation multi-center learning (CRML). First, the normalization within the class embedding multi-center learning loss effectively characterizes the direction of the representation vectors. This prevents trivial minimization of the model loss, which would otherwise degrade the separation between known and unknown classes in the representation space and impair the model's OSR performance. On the other hand, to ensure the separation between known and unknown classes in the representation space, the class embedding multi-center learning loss enhances the global directional variation of the representations. This global effect allows the strong class discrimination capability learned from known classes to be extended to both known and unknown classes, thereby improving separation between them. On both the clean and compressed sets of ADA, the CRML method demonstrated significant improvements over benchmark performance, indicating the effectiveness of CRML.

The main contributions of this paper are as follows:

\begin{itemize}
\item We introduce a novel task of audio deepfake attribution, which aims to attribute audio generated by different deepfake technologies to their sources. This work focuses on the attribution of audio generation tools.
\end{itemize}
\begin{itemize}
\item We present the first publicly available deepfake audio dataset named ADA, which is specifically focused on the attribution of audio generation tools.
\end{itemize}
\begin{itemize}
\item We introduce the novel concept of ``system fingerprints" and conduct an extensive investigation into the attribution of audio generation tools based on these fingerprints.
\end{itemize}
\begin{itemize}
\item We propose the Class-Representation Multi-Center Learning (CRML) method to address the attribution of unknown audio generation technologies, effectively reducing the open-set risks associated with model deployment in the real world.
\end{itemize}

The rest of this paper is organized as follows. 
Section II provides a review of the related work. Section III presents the design of the dataset. Section IV introduces the innovative method we proposed for open-set audio deepfake attribution. Section V reports the experimental setup and results. Section VI discusses the limitations of the current work and outlines directions for future research. Finally, the conclusions are presented.

\begin{table*}[ht]
\renewcommand\arraystretch{1.2}
\centering
\caption{Numbers of utterances, genders, speakers, and hours for training, development, and test sets are provided. The number of sentences in the clean set and the compressed set is comparable.}
\large
\resizebox{\linewidth}{!}{
\begin{tabular}{ccccccccccc}
\toprule
 \multicolumn{2}{c}{} & \multicolumn{6}{c}{Known} & \multicolumn{2}{c}{Unknown} & \multirow{2}{*}{Total} \\ \cline{3-7}  \cline{9-10}
 \multicolumn{2}{c}{} & Aispeech & Alibaba Cloud & Baidu Ai Cloud & Databaker & Sogou & Real & Tencent & iFLYTEK \\ \hline
\multirow{4}{*}{Train} 
~ & \#Utterances & 26,026 & 31,062 & 7,014 & 10,020 & 10,009 & 12,800  & -- & --  & 96,931\\
 & \#Genders & 2 & 2 & 2 & 2 & 2 & 2 & -- & -- & 2 \\ 
 & \#Speakers & 26 & 31 & 7 & 10 & 10 & 407 & -- & -- & 491 \\ 
 & \#Hours & 120.92 & 95.11 & 10.44 & 13.86 & 44.73 & 19.50 & -- & -- & 304.56 \\ \hline
\multirow{4}{*}{Dev.} 
 & \#Utterances & 8,007 & 10,020 & 2,004 & 2,004 & 4,004 & 4,800 & -- & -- & 30,839 \\ 
 & \#Genders & 2 & 2 & 2 & 2 & 2 & 2 & -- & -- & 2 \\ 
 & \#Speakers & 8 & 10 & 2 & 2 & 4 & 77 & -- & -- & 103 \\ 
 & \#Hours & 35.10 & 30.67 & 3.00 & 2.77 & 19.76 & 7.17 & -- & -- & 98.47 \\ \hline
\multirow{4}{*}{Test} 
& \#Utterances & 11,011 & 7,014 & 2,004 & 4,008 & 2,002 & 4,800 & 11,251 & 11,904 & 53,994 \\ 
 & \#Genders & 2 & 2 & 2 & 2 & 2 & 2 & 2 & 2 & 2 \\ 
 & \#Speakers & 11 & 7 & 2 & 4 & 2 & 101 & 12 & 12 & 151 \\ 
 & \#Hours & 52.51 & 21.46 & 2.76 & 5.98 & 8.99 & 7.39 & 11.56 & 12.85 & 123.5 \\ 
 \bottomrule
\end{tabular}}
\end{table*}

\section{Related Work}

Audio deepfake attribution includes identifying the audio generation algorithms and the audio generation tools. In 2022, we constructed a dataset of fake audio generated by eight algorithms, including STRAIGHT, LPCNet, WaveNet, Parallel WaveGAN, HifiGAN, Multiband-MelGAN, Style-MelGAN, and Griffin-Lim \cite{ref78}. We used vocoder fingerprints to differentiate between different generation algorithms, which laid the foundation for understanding the nuances of various TTS technologies.

Not only that, but Salvi et al. \cite{ref79} focused on the attribution of synthetic speech from various synthesis algorithms (named A01, A02, ..., A19). The experiments not only included testing the models in closed-set scenarios but also considered open-set situations, proposing two different solutions: open-set - threshold and open-set - one-class SVM to handle unknown classes.

Subsequently, we focus on analyzing speech synthesis model fingerprints in generated speech waveforms, emphasizing the roles of the acoustic model and vocoder \cite{ref80}. We extracted fingerprints from acoustic models and vocoders to identify the sources of synthesized speech. Our experiments reveal two key insights: (1) both vocoders and acoustic models leave distinct, model-specific fingerprints on generated waveforms, and (2) vocoder fingerprints, being more dominant, may obscure those from the acoustic model.

Thus, we acknowledge the limitations inherent in relying solely on vocoder fingerprints for audio deepfake attribution. Although vocoder fingerprints can be highly distinctive, reliance on the fingerprints of a single component may obscure critical features from the acoustic model or other components, thereby failing to fully differentiate between various synthesis tools or systems. Currently, advanced TTS tools from various vendors have been widely adopted. These diverse tools are likely to generate unique ``system fingerprint", making the attribution of these generation tools both necessary and meaningful. In 2023, a portion of the ADA dataset and associated baselines were employed in Track 3 of the Second Audio Deepfake Detection Challenge (ADD 2023).

\begin{figure*}
\centering
\includegraphics[height=10cm, width=\textwidth]{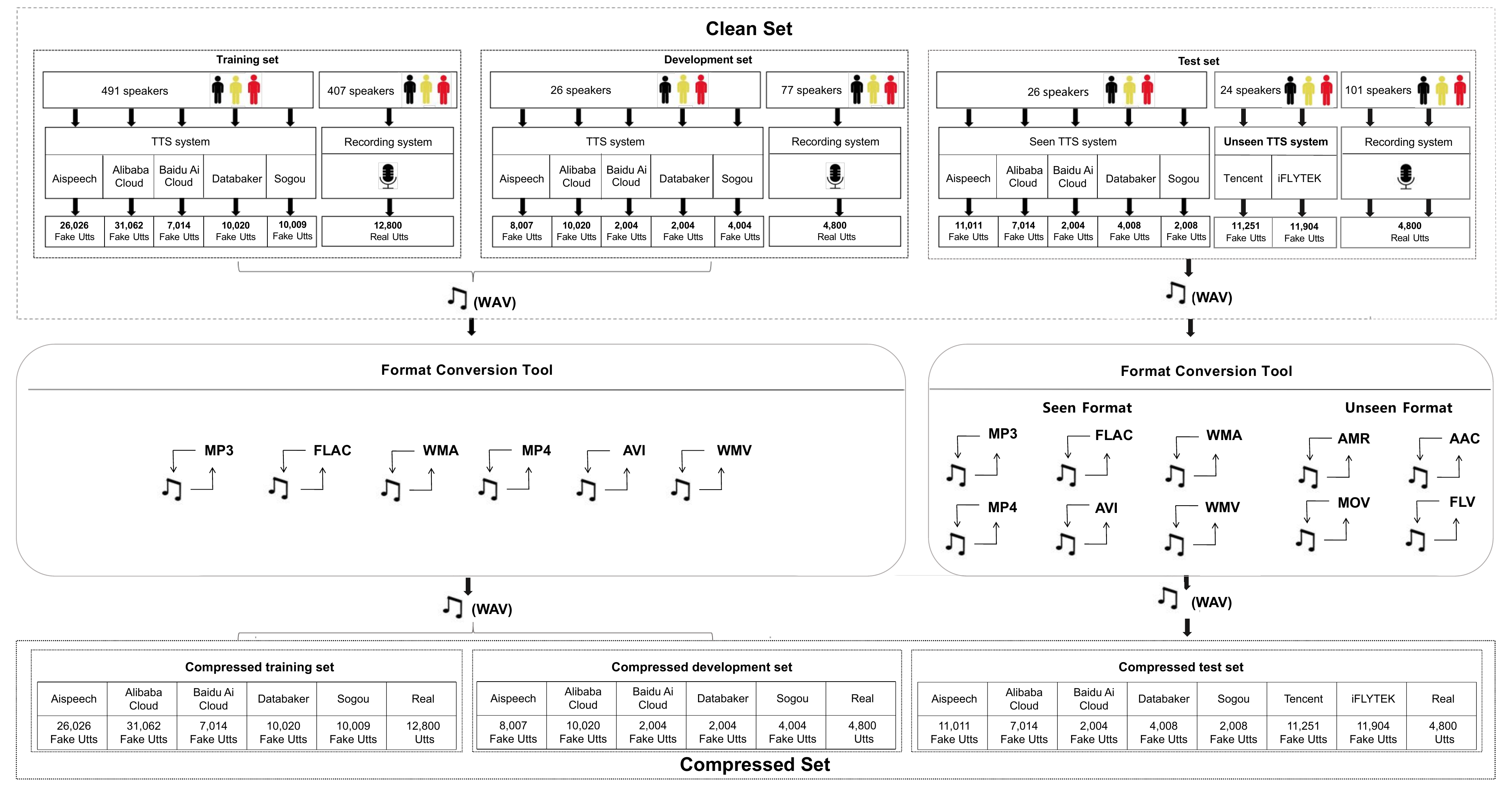}
\caption{Partitioning and construction of the SFR dataset. The top section shows the partitions of the clean set, as well as the number of speakers and the number of utterances contained in each set. The middle section simulates the data compression scenario of the data propagation process in the real world, and the bottom section shows the partitioning of the compressed set. Both
the clean set and the compressed set contain roughly equal numbers of utterances.}
\end{figure*}

\section{Dataset Design}
In order to support further research of audio deepfake attribution, this paper designs an initial dataset ADA, covering large-scale deepfake audio labelled by various Chinese mainstream TTS vendors that use the latest state-of-the-art deep learning technologies. Summarized statistics of the ADA set are shown in Table II.

\subsection{Design Policy}
We choose to collect deepfake audio from mainstream commercial application programming interfaces (APIs) and include API labels accordingly. Given that real audio can originate from various sources, the "real" label encompasses audio from multiple open-source datasets. In addition, The main characteristic of deepfake is the transmission on social media. They are inevitably subject to the data compression problems associated with transcoding. Hence we conduct the following work separately (1) deepfake audio collection, (2) real audio collection, (3) data compression.

\subsection{Deepfake Audio Collection}
This part presents the core portion of the dataset, which is the deepfake audio section. We selected seven mainstream open-source vendors, designating five as known generative tools and the remaining two as unknown generative tools.

The first problem we faced was the selection of the input corpus. The input to the TTS system required a large number of different phrases to ensure that the underlying distribution was well represented in the dataset. In addition, considering the design goals of our dataset, we tried to find a corpus that matches the characteristics of the task. For these two reasons, we chose the Chinese text dataset called THUCNews\footnote{http://thuctc.thunlp.org/message}. HUCNews includes 740,000 news documents, all in plain text format. Once the corpus is ready, we use the APIs of each TTS system to generate speech segments.

Each generative tool utilizes the latest advanced TTS technologies developed by their respective platforms. The audio generated by known generative tools comes from Aispeech\footnote{https://cloud.aispeech.com/openSource/technology/tts}, Sogou\footnote{https://ai.sogou.com}, Alibaba Cloud\footnote{https://ai.aliyun.com/nls/tts}, Baidu Ai Cloud\footnote{https://ai.baidu.com/tech/speech/tts}, and Databaker\footnote{https://www.databaker.com/specs/compose/online}. This component of the dataset comprises 141,009 audio samples from 136 speakers.

Additionally, the open set used for open-set audio deepfake Attribution includes audio generated by two unknown generative tools: Tencent Cloud\footnote{https://cloud.tencent.com/product/tts} and iFLYTEK\footnote{https://www.xfyun.cn/services/online\_tts}. Samples from these unknown generative tools are only present in the testing phase and are categorized as a single unknown (UNK) class. The open set ultimately includes 23,155 audio samples from 24 speakers.

\subsection{Real Audio Collection}
The real audio data was selected from four open-source datasets \cite{ref32} \cite{ref33} \cite{ref34} \cite{ref35}:

AISHELL-1\footnote{http://openslr.org/33/}: The audio of the dataset is derived from recordings of 400 people from different accent regions of China, recorded in a quiet indoor environment. 

AISHELL-3\footnote{http://openslr.org/93/}: The dataset contains 88,035 recordings by 218 Mandarin speakers. 

THCHS-30\footnote{http://openslr.org/18/}: THCHS-30 consists of over 30 hours of speech recorded in a quiet environment. 

Aidatatang\_200zh\footnote{https://openslr.org/62/}: 
A corpus containing 200 hours of speech data from 600 speakers. 

Finally a total of 22,400 audios were selected as the real benchmark dataset.

\subsection{Data Compression}
In the real world, the system fingerprints will experience different scenarios. Due to the spread of multimedia content posted online, audio media is inevitably affected by data compression, as shown in Fig 2. This is evident in scenarios such as incorporating audio into secondary video creations or disseminating audio across various social media platforms. In addition, the model will encounter unseen samples in the real world. So we transcode between audio and audio, and between audio and video, to simulate the changes of the system fingerprints propagated in realistic scenarios. We created a compressed set based on the clean set. To face open-world challenges, one part of the test set was processed with unseen compression. We used a professional tool called Golden Lion Video Assistant to complete the transcoding operation. The build process for the compressed set is as follows: 

The clean set is processed with a set of codecs giving the ten compression conditions shown in Table III. The conditions Audio-C1 - Audio-C5 use diffirent audio codecs. Conditions Video-C1 - Video-C5 use different video codecs. The difference between the same codec conditions is the use of different variable bit rate (VBR) configurations, from low to high, as shown in the rightmost column of Table III. The compressed audio data and video data are transcoded into WAV format with 16 kHz sample rate and 256 kbps bit rate to compose the compressed set.

\begin{table}[!t]
\renewcommand\arraystretch{1.2}
\caption{Summary of data compression conditions. \underline{The underlined factors} appear only in the test subset.\label{tab:table1}}
\small
\centering
\begin{tabular}{cccc}
\toprule
Cond.& Codec & Compression & Bitrate\\
\hline
Audio-C1 & MP3 & MP3 & 128 kbps\\

Audio-C2 & FALC & FALC & $\sim$130-210 kbps\\

Audio-C3 & WMA & WMA & 128 kbps\\

\underline{Audio-C4} & AMR & AMR & 12 kbps\\

\underline{Audio-C5} & AAC & AAC & $\sim$60-80 kbps\\

Video-C1 & H.264 & MP4 & $\sim$90-100 kbps\\

Video-C2 & H.264 & AVI & 256 kbps\\

Video-C3 & WMV9 & WMV & 1411 kbps\\

\underline{Video-C4} & H.264 & MOV & $\sim$90-100 kbps\\

\underline{Video-C5} & H.264 & FLV & 125 kbps\\
\bottomrule
\end{tabular}
\vspace{-30pt}
\end{table}

\begin{figure*}
\centering
\includegraphics[width=\textwidth]{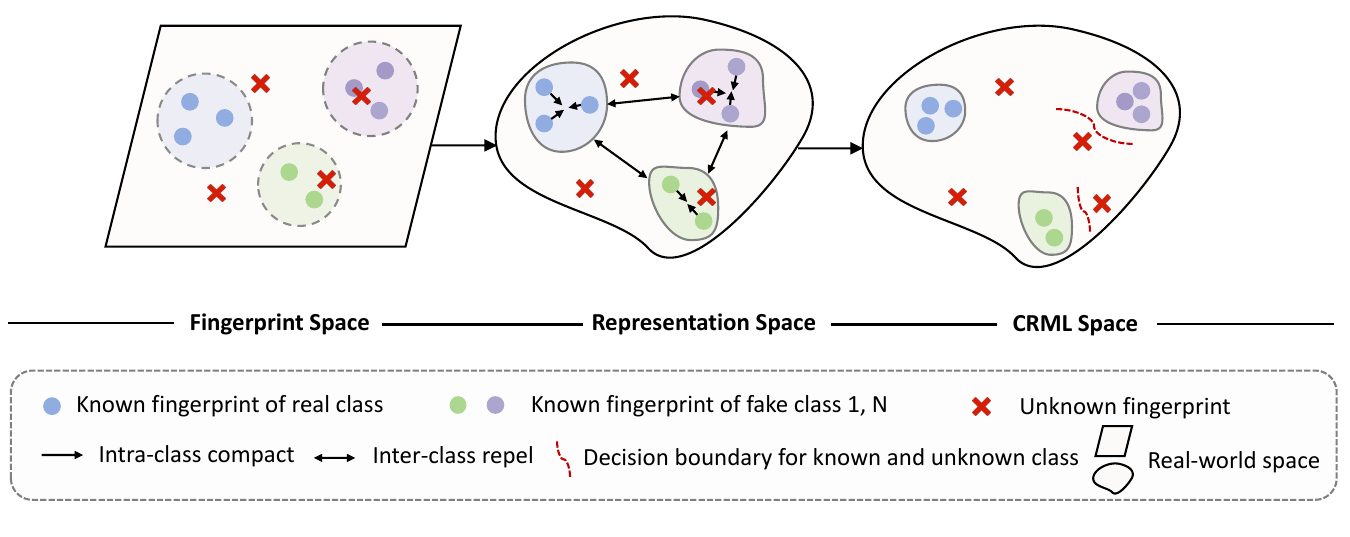}
\caption{A schematic diagram of OSADA based on class-representation multi-center learning (CRML). Firstly, CRML fosters intra-class compactness and inter-class separability within the representation space. To ensure the separation between known and unknown classes in real-world space, CRML enhances the global directional variation of representations. The robust class discriminative capability learned from known classes can effectively distinguish the relationship between known and unknown classes.}
\end{figure*}

\subsection{Dataset Composition}
The ADA dataset consists of two subsets: (1) Clean set and (2) Compressed set. Future researchers in related fields can use these two subsets to compare the performance of audio deepfake attribution methods. Both the clean set and the compressed set are divided into training (Train) set, development (Dev.) set, and testing (Test) set. The speakers of the three sets are disjoint. Both the clean set and the compressed set contain roughly equal numbers of utterances.

\section{Open-World Audio Deepfake Attribution: Attribution of Unknown Audio Generation Tools}
New classes continuously emerge in the real world, which may not have been observed during model training. Therefore, system fingerprint attribution systems need to reject unknown audio generation tools, a problem commonly referred to as open-set recognition (OSR) \cite{ref50}\cite{ref52}\cite{ref67}\cite{ref68}\cite{ref69}. To simulate real-world open scenarios, we extended the ADA test dataset by adding an open set. Additionally, we propose a method called Class-Representation Multi-Center Learning (CRML) for Open-Set Audio Deepfake Attribution (OSADA). A schematic diagram of class-representation multi-center learning in the fingerprint space is shown in Fig 3. In the following subsections, each critical part of the proposed method will be explained in detail.

\subsection{Class-Representation Multi-Center Learning (CRML)}

The OSR model based on discriminative representation learning aims to learn a robust representation that effectively separates unknown classes from known classes in the representation space. A potential solution is to obtain well-clustered representations of known class data, which exhibit strong inter-class differences and intra-class similarities. This representational distinction may transfer to the global data, thereby achieving separation between known and unknown classes in the representation space.


However, without proper regularization, the strong discriminative power for known classes does not necessarily extend to both known and unknown classes. One issue is that the size of the representations is unrestricted. Such unrestricted models can induce strong discriminative power on known classes but may perform poorly in separating known and unknown classes in the representation space. Another important aspect to consider in an OSR environment is that actual unknown class samples often lie near the boundaries of known classes. Therefore, the representation must be sensitive to input variations in the open space to generate distinct separations between known and unknown classes (Fig.3).


Since the magnitude of representations ($\| g(x) \|_2$) is a non-discriminative factor, it is removed through normalization. The normalization of the representation $g(x)$ and class prototypes $w_k$ is as follows, where these prototypes act as proxies for known classes in the representation space: 
\begin{equation}
g(x) = \frac{\hat{g}(x)}{\| \hat{g}(x) \|_2}
\end{equation}
\begin{equation}
\setlength{\belowdisplayskip}{5pt} 
\setlength{\belowdisplayshortskip}{5pt} 
w_k = \frac{\hat{w}_k}{\| \hat{w}_k \|_2}
\end{equation}


Where $\hat{g}(x)$ and $\hat{w}_k$ are the pre-normalized representations in the model, and $\| \cdot \|_2$ denotes the Euclidean norm. Therefore, normalization characterizes the representation $g(x)$ entirely through its vector direction. After normalization, the proposed loss is minimized to maximize its discriminative power for known classes.
\begin{equation}
\setlength{\abovedisplayskip}{1pt}
\setlength{\belowdisplayskip}{1pt}
L(x, y; f, t, s) = \sum_{k=1}^{K} \log \left(1 + e^{(-1)^m s \cos(\theta_k + t)}\right)
\end{equation}


Where $L(x, y) = L(x, y; f, t, s)$ depends on $f$, $t$, and $s$. Here, $\cos \theta_k = g(x) \cdot w_k$ represents the similarity between $g(x)$ and $w_k$, $f$ is a parameterized classifier such that $f_k(x) = \cos \theta_k$, and $s > 0$ is a scaling hyperparameter to accelerate training. On the other hand, $t$ is a calibration hyperparameter used to adjust the decision boundary to provide better separation between classes that are difficult to distinguish. $m$ is a binary variable that is 1 when $k$ is the real class index, and 0 otherwise. Minimizing $L$ enhances the similarity between $(g(x), w_y)$ within the same class and increases dissimilarity between $(g(x), w_k)$ across different classes. 


Finally, the network is jointly supervised by the softmax loss and the CRML, as shown below:
\begin{equation}
L = L_{\text{Softmax}} + \lambda L(x, y; f, t, s)
\end{equation}
The hyperparameter $\lambda$ here is used to balance the two loss. We summarize the training process
in Alg. 1.

\begin{algorithm}[H]
\caption{Training of CRML}\label{alg:crml_training}
\begin{algorithmic}
\REQUIRE Training data $(x, y)$; Initial neural network parameters $\theta$; Learnable Class-Representation Multi-Center $W = \{w_k\}_{k=1}^{N}$; Hyperparameters $t$, $s$, and $\lambda$; Iteration count $\text{iter} = 0$.
\ENSURE Trained parameters $\theta$ and Class-Representation Multi-Center $W$.
\STATE \textbf{procedure}
\STATE \textbf{while} training \textbf{has not converged} \textbf{do}
\STATE \hspace{0.5cm} $\text{iter} \gets \text{iter} + 1$
\STATE \hspace{0.5cm} \textbf{for} each batch $(x_{\text{batch}}, y_{\text{batch}})$ \textbf{in} $(x, y)$ \textbf{do}
\STATE \hspace{1cm} Compute Class-Representation Multi-Center Learning loss: 
\STATE \hspace{1cm} $L(x_{\text{batch}}, y_{\text{batch}}; f, t, s) = \sum_{k=1}^{K} \log \left(1 + e^{(-1)^m s \cos(\theta_k + t)} \right)$
\STATE \hspace{1cm} Compute the joint supervised loss: 
\STATE \hspace{1cm} $L = L_{Softmax} + \lambda L(x_{batch}, y_{batch}; f, t, s)$
\STATE \hspace{1cm} Update Class-Representation Multi-Center $W$
\STATE \hspace{1cm} Update parameters $\theta$
\STATE \hspace{0.5cm} \textbf{end for}
\STATE \textbf{end while}
\end{algorithmic}
\label{alg1}
\end{algorithm}

\subsection{Inference}
Even if there is a reasonable separation between the representations of known and unknown classes, an OSR model still requires a metric to capture this potential separation effectively. To achieve this, for a query $x$, the inference score $S(x)$ is defined as the pseudo-model loss using the proposed loss function:

\begin{equation}
\label{eqn_inference}
S(x) = L(x, p; f, 0, s)
\end{equation}

\begin{equation}
\label{eqn_p}
p = \arg \max_{k} \cos(\theta_k)
\end{equation}

Here, the pseudo-label \( p \) is the value of \( k \) that maximizes \( \cos(\theta_k) \), with the correction hyperparameter \( t = 0 \).

The proposed inference score is a decreasing function that depends on the similarity between the representation $g(x)$ and the nearest known class prototype $w_p$. At the same time, it is an increasing function that depends on the similarity between $g(x)$ and other prototypes $w_k$ ($k \ne p$). Therefore, if $g(x)$ is close to any known class prototype $w_k$, it is close to $w_p$ and far from $w_k$ ($k = p$). In this case, $S(x)$ will be small, and $x$ will be detected as a known class. Otherwise, $x$ will be detected as an unknown class. Overall, the category of the query $x$ is determined by

\begin{equation}
\label{eq:piecewise}
y^* =
\begin{cases}
\text{unknown} & \text{if } S(x) > \tau \\
\text{p-th class} & \text{otherwise}
\end{cases}
\end{equation}
for a threshold $\tau > 0$.

\section{Experiments}
In this section, we first present extensive benchmark experiments on system fingerprint recognition and evaluate both mainstream pipeline and end-to-end models. Then, we propose a method called class-representation multi-center learning (CRML) for open-set audio deepfake attribution (OSADA). Finally, we visualize the system fingerprints to better explain the effectiveness of the audio deepfake attribution.


\subsection{Evaluation Metrics}
\begin{itemize}
\item Precision (P): measures the proportion of predictions made by the model for a particular class that are actually correct.

\item Recall (R): measures the proportion of all actual positive samples in a particular class that the model correctly predicts.

\item F\textsubscript{1}-score (F\textsubscript{1}): is the harmonic mean of precision and recall for each class.

\item Accuracy (Acc.): measures the model's overall ability to classify samples across all categories correctly. 

\item In-Distribution Accuracy (ID Acc.): measures whether the OSR  method can maintain robust classification performance for known categories while rejecting unknown classes.

\end{itemize}


\subsection{Experimental Setup}
\textbf{Feature extraction:} 

In our baseline experiments, the front-end features include hand-crafted features and self-supervised embedding features.
\begin{itemize}
\item Constant Q Cepstral Coefficients (CQCC) \cite{ref37}: Long-term spectral features are extracted according to the system \footnote{https://github.com/asvspoof-challenge/2021/tree/main/LA/Baseline-CQCC-GMM}.

\item Mel-Frequency Cepstral Coefficients (MFCC) \cite{ref38}: The Hamming window analysis is performed with a window size of 25 ms and a shift of 10 ms.

\item Linear Frequency Cepstral Coefficients (LFCC) \cite{ref38}: Short-term spectral features are extracted following the configuration detailed in the system \footnote{https://github.com/asvspoof-challenge/2021/tree/main/LA/Baseline-LFCC-LCNN}.

\item Wav2vec 2.0 \cite{ref70}: Features are extracted from the self-supervised Wav2Vec 2.0 pre-trained model \footnote{https://github.com/facebookresearch/fairseq/tree/main/examples/wav2vec}.

\item Wav2vec 2.0 XLS-R \cite{ref71}: We evaluate the large-scale, cross-lingually pre-trained model \footnote{https://github.com/facebookresearch/fairseq/tree/main/examples/wav2vec/xlsr} in our benchmarks.

\item HuBERT \cite{ref72}: Speech feature extraction is enhanced by predicting masked portions of the audio \footnote{https://github.com/facebookresearch/fairseq/tree/main/examples/hubert}.

\item WavLM \cite{ref73}: We utilize a large-scale, pre-trained self-supervised model to derive high-quality speech features \footnote{https://github.com/microsoft/unilm/tree/master/wavlm}.
\end{itemize}

\textbf{Model architecture:} An effective classification model plays an important role. The baseline models include pipeline and end-to-end architectures, which are mainstream in fake audio detection or voiceprint recognition
\begin{itemize}
\item X-vector \cite{ref40}: We use the x-vector system built on the TDNN embedding architecture.

\item SE-ResNet \cite{ref42}: Each SE-ResNet module comprises two convolutional layers with 3 × 3 kernels and one SE block.

\item ResNet \cite{ref43}: We choose an 18-layer ResNet that converges faster.

\item RawNet2 \cite{ref44}: We choose the RawNet2 baseline of the ASVspoof 2021, which operates on the raw speech waveform.

\item RawGAT-ST \cite{ref75}: This is an end-to-end model that leverages graph attention mechanisms for sequence-to-sequence tasks in raw audio processing.

\item AASIST \cite{ref61}: This is an end-to-end model designed for acoustic scene analysis using self-supervised learning techniques.

\item wav2vec2.0-AASIST \cite{ref76}: This is an end-to-end model that integrates wav2vec 2.0 features with AASIST.

\item RawFormer \cite{ref77}: This is an end-to-end model that applies transformer-based architectures to raw audio for enhanced speech representation.

\item RawBMamba \cite{ref66}: This is an end-to-end model that combines raw audio processing with advanced neural network techniques.
\color{black} 
\end{itemize}
\textbf{Training strategy:} Parameters are randomly initialized. The training was conducted with the Adam optimizer to accelerate optimization by applying an adaptive learning rate. The initial learning rate is set to 0.001 for Adam, with linear learning rate decay. These models are trained with a mini-batch size of 128 for 100 epochs.

\begin{table*}[ht]
\renewcommand\arraystretch{1.2}
\small
\centering
\caption{Comparison of system fingerprint attribution performance with different handcrafted features. The best performance is highlighted in bold.}
\resizebox{0.8\textwidth}{!}{
\begin{tabular}{ccccccccccc}
\toprule
\multicolumn{2}{c}{\multirow{2}{*}{Model}}  & \multicolumn{3}{c}{Clean Set} & \multicolumn{3}{c}{Compressed Set} & \multicolumn{3}{c}{Average} \\  
\cmidrule(lr){3-5} \cmidrule(lr){6-8} \cmidrule(lr){9-11}
\multicolumn{2}{c}{} & \multicolumn{1}{c}{P $\uparrow$} & \multicolumn{1}{c}{R $\uparrow$} & F\textsubscript{1} $\uparrow$      & \multicolumn{1}{c}{P $\uparrow$} & \multicolumn{1}{c}{R $\uparrow$} & F\textsubscript{1} $\uparrow$ & \multicolumn{1}{c}{P $\uparrow$} & \multicolumn{1}{c}{R $\uparrow$} & \multicolumn{1}{c}{F\textsubscript{1}} \\  
\hline 
\multicolumn{1}{c}{\multirow{4}{*}{CQCC}} & X-vector & \multicolumn{1}{c}{77.00}     & \multicolumn{1}{c}{76.07}  & 74.08 & \multicolumn{1}{c}{62.73}     & \multicolumn{1}{c}{60.32}  & 60.85 & \multicolumn{1}{c}{69.87} & \multicolumn{1}{c}{68.20} & 67.47 \\ 
\multicolumn{1}{c}{}                      & LCNN     & \multicolumn{1}{c}{83.32}     & \multicolumn{1}{c}{83.15}  & 82.81 & \multicolumn{1}{c}{66.57}     & \multicolumn{1}{c}{67.98}  & 66.74 & \multicolumn{1}{c}{74.95} & \multicolumn{1}{c}{75.57} & 74.77 \\  
\multicolumn{1}{c}{}                      & SENet    & \multicolumn{1}{c}{85.79}     & \multicolumn{1}{c}{85.98}  & 85.82 & \multicolumn{1}{c}{74.15}     & \multicolumn{1}{c}{70.63}  & 69.98 & \multicolumn{1}{c}{79.97} & \multicolumn{1}{c}{78.81} & 77.90 \\  
\multicolumn{1}{c}{}                      & ResNet   & \multicolumn{1}{c}{93.19}     & \multicolumn{1}{c}{94.45}  & 93.71 & \multicolumn{1}{c}{77.88}     & \multicolumn{1}{c}{76.83}  & 76.14 & \multicolumn{1}{c}{85.54} & \multicolumn{1}{c}{85.64} & 84.93 \\ \hline
\multicolumn{1}{c}{\multirow{4}{*}{MFCC}} & X-vector & \multicolumn{1}{c}{81.23}     & \multicolumn{1}{c}{79.57}  & 80.07 & \multicolumn{1}{c}{71.86}     & \multicolumn{1}{c}{72.39}  & 71.22 & \multicolumn{1}{c}{76.55} & \multicolumn{1}{c}{75.98} & 75.65 \\ 
\multicolumn{1}{c}{}                      & LCNN     & \multicolumn{1}{c}{87.61}     & \multicolumn{1}{c}{85.88}  & 86.51 & \multicolumn{1}{c}{75.51}     & \multicolumn{1}{c}{74.81}  & 74.56 & \multicolumn{1}{c}{81.56} & \multicolumn{1}{c}{80.35} & 80.53 \\ 
\multicolumn{1}{c}{}                      & SENet    & \multicolumn{1}{c}{91.89}     & \multicolumn{1}{c}{90.60}  & 90.99 & \multicolumn{1}{c}{\textbf{79.23}}     & \multicolumn{1}{c}{\textbf{77.79}}  & \textbf{77.79} & \multicolumn{1}{c}{85.56} & \multicolumn{1}{c}{84.19} & 84.89 \\ 
\multicolumn{1}{c}{}                      & ResNet   & \multicolumn{1}{c}{92.84}     & \multicolumn{1}{c}{89.88}  & 91.05 & \multicolumn{1}{c}{77.94}     & \multicolumn{1}{c}{75.75}  & 76.25 & \multicolumn{1}{c}{85.89} & \multicolumn{1}{c}{82.82} & 83.15 \\ \hline
\multicolumn{1}{c}{\multirow{4}{*}{LFCC}} & X-vector & \multicolumn{1}{c}{86.81}     & \multicolumn{1}{c}{86.27}  & 86.29 & \multicolumn{1}{c}{68.84}     & \multicolumn{1}{c}{74.05}  & 70.99 & \multicolumn{1}{c}{77.83} & \multicolumn{1}{c}{80.16} & 78.64 \\ 
\multicolumn{1}{c}{}                      & LCNN     & \multicolumn{1}{c}{94.92}     & \multicolumn{1}{c}{96.11}  & 95.27 & \multicolumn{1}{c}{72.15}     & \multicolumn{1}{c}{74.47}  & 72.85 & \multicolumn{1}{c}{83.54} & \multicolumn{1}{c}{85.29} & 84.06 \\ 
\multicolumn{1}{c}{}                      & SENet    & \multicolumn{1}{c}{96.26}     & \multicolumn{1}{c}{96.83}  & 96.51 & \multicolumn{1}{c}{77.43}     & \multicolumn{1}{c}{75.81}  & 75.87 & \multicolumn{1}{c}{86.85} & \multicolumn{1}{c}{86.32} &86.19 \\ 
\multicolumn{1}{c}{}                      & ResNet   & \multicolumn{1}{c}{\textbf{98.95}}     & \multicolumn{1}{c}{\textbf{99.00}}  & \textbf{98.97} & \multicolumn{1}{c}{78.56}     & \multicolumn{1}{c}{76.12}  & 77.34 & \multicolumn{1}{c}{\textbf{88.75}} & \multicolumn{1}{c}{\textbf{87.56}} & \textbf{87.81} \\ \bottomrule
\end{tabular}
}
\end{table*}

\begin{table*}
\renewcommand\arraystretch{1.2}
\small
\centering
\caption{Comparison of system fingerprint attribution performance with different self-supervised embedding features. The best performance is displayed in bold.}
\resizebox{0.9\textwidth}{!}{
\begin{tabular}{ccccccccccc}
\toprule
\multicolumn{2}{c}{\multirow{2}{*}{Model}}  & \multicolumn{3}{c}{Clean Set} & \multicolumn{3}{c}{Compressed Set} & \multicolumn{3}{c}{Average} \\  
\cmidrule(lr){3-5} \cmidrule(lr){6-8} \cmidrule(lr){9-11}
\multicolumn{2}{c}{} & \multicolumn{1}{c}{P $\uparrow$} & \multicolumn{1}{c}{R $\uparrow$} & F\textsubscript{1} $\uparrow$      & \multicolumn{1}{c}{P $\uparrow$} & \multicolumn{1}{c}{R $\uparrow$} & F\textsubscript{1} $\uparrow$ & \multicolumn{1}{c}{P $\uparrow$} & \multicolumn{1}{c}{R $\uparrow$} & F\textsubscript{1} $\uparrow$ \\  
\hline
\multicolumn{1}{c}{\multirow{4}{*}{wav2vec 2.0}} & X-vector & 81.65     & 81.78  & 78.39 & 70.01     & 65.29  & 60.28 & 75.83 & 73.53 & 69.34 \\  
\multicolumn{1}{c}{}                      & LCNN     & 94.88     & 94.72  & 94.67  & 74.11     & 76.65  & 74.96 & 84.49 & 85.69 & 84.82 \\  
\multicolumn{1}{c}{}                      & SENet    & 98.62     & 98.61  & 98.61 & 77.14     & 77.98  & 77.01 & 87.88 & 88.30 & 87.81 \\  
\multicolumn{1}{c}{}                      & ResNet   & \multicolumn{1}{c}{98.51}     & \multicolumn{1}{c}{98.50}  & 98.50 & \multicolumn{1}{c}{57.48}     & \multicolumn{1}{c}{72.26}  & 63.74 & 78.00 & 85.38 & 81.12 \\ \hline
\multicolumn{1}{c}{\multirow{4}{*}{wav2vec 2.0 XLS-R}} & X-vector  & 60.63 & 49.09 & 47.66  & 72.48     & 50.74  & 38.70  & 66.56 & 49.91 & 43.68 \\  
\multicolumn{1}{c}{}                      & LCNN     & 90.20     & 89.23  & 89.25 & 74.34     & 72.01  & 72.52 & 82.27 & 80.12 & 80.38 \\  
\multicolumn{1}{c}{}                      & SENet    & 98.67     & 98.65  & 98.65 & \textbf{91.05}     & \textbf{90.76}  & \textbf{90.47} & \textbf{94.36} & \textbf{94.71} & \textbf{94.56} \\  
\multicolumn{1}{c}{}                      & ResNet   & 97.39     & 97.35  & 97.35  & 79.90     & 84.62  & 82.09 & 88.64 & 90.99 & 89.72 \\ \hline
\multicolumn{1}{c}{\multirow{4}{*}{HuBERT}} & X-vector & 93.94     & 93.39  & 92.88 & 69.87     & 69.02  & 62.23 & 81.91 & 81.21 & 77.55 \\  
\multicolumn{1}{c}{}                      & LCNN     & \multicolumn{1}{c}{98.80}     & \multicolumn{1}{c}{98.80}  & 98.80 & \multicolumn{1}{c}{87.33}     & \multicolumn{1}{c}{87.03}  & 86.67 & 93.06 & 92.92 & 92.74 \\  
\multicolumn{1}{c}{}                      & SENet    & 98.52     & 98.48  & 98.46 & 83.76     & 85.80  & 83.42 & 91.14 & 92.14 & 90.94 \\  
\multicolumn{1}{c}{}                      & ResNet   & 98.87     & 98.86  & 98.86 & 84.27     & 86.03  & 84.92 & 91.57 & 92.45 & 90.89 \\ \hline
\multicolumn{1}{c}{\multirow{4}{*}{WavLM}}& X-vector & 91.98     & 91.42  & 90.76 & 79.07     & 67.43  & 63.35 & 85.53 & 79.43 & 77.05 \\  
\multicolumn{1}{c}{}                      & LCNN     & 98.13     & 98.12  & 98.11 & 82.73     & 82.60  & 81.51 & 90.43 & 90.36 & 89.81 \\  
\multicolumn{1}{c}{}                      & SENet    & \textbf{99.10} & \textbf{99.10}  & \textbf{99.10} & 89.35     & 89.64  & 89.39 & 94.22 & 94.37 & 94.25 \\  
\multicolumn{1}{c}{}                      & ResNet   & 98.72     & 98.72  & 98.72 & 80.13     & 83.87  & 81.63 & 88.43 & 91.29 & 89.18 \\ \bottomrule
\end{tabular}}
\end{table*}

\begin{table*}[ht]
\renewcommand\arraystretch{1.2}
\small
\centering
\caption{Comparison of system fingerprint attribution performance with different end-to-end models. The best performance is displayed in bold.}
\resizebox{0.8\textwidth}{!}{
\begin{tabular}{cccccccccc}
\toprule
\multirow{2}{*}{Model} & \multicolumn{3}{c}{Clean Set} & \multicolumn{3}{c}{Compressed Set} & \multicolumn{3}{c}{Average} \\  
\cmidrule(lr){2-4} \cmidrule(lr){5-7} \cmidrule(lr){8-10}
 & \multicolumn{1}{c}{P $\uparrow$} & \multicolumn{1}{c}{R $\uparrow$} & F\textsubscript{1} $\uparrow$      & \multicolumn{1}{c}{P $\uparrow$} & \multicolumn{1}{c}{R $\uparrow$} & F\textsubscript{1} $\uparrow$ & \multicolumn{1}{c}{P $\uparrow$} & \multicolumn{1}{c}{R $\uparrow$} & F\textsubscript{1} $\uparrow$ \\  
\hline
RawNet2 & 85.10 & 82.63 & 83.50 & 82.65 & 78.74 & 79.64 & 83.38 & 80.68 & 81.57 \\  
RawGAT-ST & 84.67 & 83.85 & 80.70 & 48.63 & 50.99 & 49.31 & 66.65 & 67.42 & 65.01 \\  
AASIST & 74.36 & 74.67 & 73.07 & 54.21 & 49.26 & 50.42 & 64.29 & 61.97 & 61.75 \\  
wav2vec2.0-AASIST & 81.06 & 79.68 & 79.19 & 56.35 & 51.44 & 50.57 & 68.71 & 65.56 & 64.88 \\  
RawFormer & 93.05 & 92.29 & 92.00 & \textbf{85.29} & \textbf{83.45} & \textbf{82.62} & \textbf{89.17} & \textbf{87.87} & \textbf{87.31} \\
RawBMamba & \textbf{93.21} & \textbf{93.31} & \textbf{93.13} & 82.67 & 81.90 & 81.12 & 87.94 & 87.61 & 87.13 \\ 
\bottomrule
\end{tabular}
}
\label{tab:end-to-end-models}
\end{table*}

\begin{table*}[ht]
\renewcommand\arraystretch{1}
\caption{Cross-dataset generalization. Comparison of system fingerprint attribution performance when training on one dataset and testing on another.}
\small
\centering
\begin{tabularx}{\textwidth}{c X X X X X X X X X X X}
\toprule
 \multirow{2}{*}{Subset} & \multirow{2}{*}{Train} & \multirow{2}{*}{Test} & \multicolumn{3}{c}{WavLM-SENet} & \multicolumn{3}{c}{wav2vec2.0 XLS-R-SENet} & \multicolumn{3}{c}{Average} \\ 
\cmidrule(lr){4-6} \cmidrule(lr){7-9} \cmidrule(lr){10-12}
 & & & \multicolumn{1}{c}{P $\uparrow$} & \multicolumn{1}{c}{R $\uparrow$} &{F\textsubscript{1}}$\uparrow$ & \multicolumn{1}{c}{P $\uparrow$} & \multicolumn{1}{c}{R $\uparrow$} & F\textsubscript{1} $\uparrow$& \multicolumn{1}{c}{P $\uparrow$} & \multicolumn{1}{c}{R $\uparrow$} & F\textsubscript{1} $\uparrow$\\
\midrule
Clean Set & \checkmark & &\multirow{2}{*}{87.08}  & \multirow{2}{*}{75.46} &\multirow{2}{*}{77.96}   &\multirow{2}{*}{85.71}  & \multirow{2}{*}{77.31} &\multirow{2}{*}{78.72}   & \multirow{2}{*}{86.40} &  \multirow{2}{*}{76.39} & \multirow{2}{*}{78.34}\\
Compressed Set & & \checkmark &  &  & &   &  &  & \\
\hline
Clean Set & & \checkmark &  \multirow{2}{*}{94.10}  & \multirow{2}{*}{94.14}  & \multirow{2}{*}{94.03} &\multirow{2}{*}{94.62}  &\multirow{2}{*}{94.23}   & \multirow{2}{*}{94.00}&  \multirow{2}{*}{94.36} &  \multirow{2}{*}{ 94.19} &   \multirow{2}{*}{94.02}\\
Compressed Set & \checkmark & &  &  & &  &  & &  &  & \\
\bottomrule
\end{tabularx}
\end{table*}

\begin{table*}[ht]
\renewcommand\arraystretch{1.3}
\small
\centering
\caption{Open-set audio deepfake attribution for audio generation tools. Comparison of the performance of our CRML method with classical baseline methods. The best performance is indicated in bold.}
\begin{tabularx}{\textwidth}{c X X X X X X X X X X}
\toprule
 \multirow{2}{*}{Method} & \multicolumn{3}{c}{Clean Set} & \multicolumn{3}{c}{Compressed Set} & \multicolumn{3}{c}{Average} \\
\cmidrule(lr){2-4} \cmidrule(lr){5-7} \cmidrule(lr){8-10}
& F\textsubscript{1} $\uparrow$ & Acc. $\uparrow$ & ID Acc. $\uparrow$ & F\textsubscript{1} $\uparrow$ & Acc. $\uparrow$ & ID Acc. $\uparrow$ & F\textsubscript{1} $\uparrow$  & Acc. $\uparrow$ & ID Acc. $\uparrow$ \\
\midrule
SoftMax & 72.82 & 60.45 & \textbf{97.82} & 54.87 & 51.92 & 79.71 & 63.85 & 56.19 & 88.77 \\
OpenMax \cite{ref53} & 88.06 & 67.04 & 97.57 & 58.82 & 57.73 & 78.62 & 73.44 & 62.39 & 88.10 \\
CROSR \cite{ref51} & 88.94 & 69.23 & 96.47 & 62.38 & 66.59 & 79.95 & 75.66 & 67.91 & 88.21 \\
\hline
\textbf{CRML (Ours)} & \textbf{92.35} & \textbf{87.25} & 96.98 & \textbf{69.34} & \textbf{67.95} & \textbf{82.55} & \textbf{80.85} & \textbf{77.60} & \textbf{89.77} \\
\bottomrule
\end{tabularx}
\end{table*}

\subsection{Which features and models can better distinguish the attribution of system fingerprints?}
We conducted extensive experiments to compare system fingerprint attribution performance, including advanced pipeline models (Tables IV and V) and end-to-end models (Table VI).

For pipeline models, the results show that WavLM-SENet can better distinguish the attribution of system fingerprints in clean environments. This model achieved an F\textsubscript{1} score of 99.10\%, the highest among all models. Additionally, wav2vec 2.0 XLS-R-SENet has a significant performance advantage on the compressed set, demonstrating its adaptability to compressed environments.

Overall, self-supervised features generally outperform handcrafted features, which may be attributed to their ability to learn richer speech representations from large-scale data. However, in cases of lower data quality (such as compressed data), handcrafted features show stronger stability with less performance degradation. Most models perform well in distinguishing the attribution of system fingerprints in clean environments, especially SENet. However, the performance of models generally declines on compressed sets.

For end-to-end models, the results indicate that RawBMamba achieves the best performance on the clean set and also demonstrates certain advantages on the compressed set. All models exhibit generally poorer performance after data compression, suggesting that future design of system fingerprint attribution models should account for adaptability to data compression.

\subsection{Cross-dataset generalization. }

The effectiveness obtained by a method on a given dataset
do not always reflect its robustness in practical applications. Therefore, we evaluate how system fingerprint attribution methods react to interchanging the Train set of dataset X by the Train set of dataset Y and test it on the Test set of X. X and Y refer to one of the clean and compressed sets, respectively. This is referred to as evaluating cross-dataset generalization in this paper. The experimental results are shown in Table VII. After analyzing the experimental results presented in Table V and Table VII, it is observed that, irrespective of the model architecture, training on the clean set and testing on the compressed set results in performance degradation. Conversely, training with the compressed set typically leads to improved performance, with this benefit becoming more pronounced when evaluating on the clean set.

\subsection{Open-Set Audio Deepfake Attribution (OSADA)}

The OSADA task involves detecting unknown class samples in an open set setting, as well as attributing known class samples in a closed set setting. CRML is compared with classic baselines for OSR, including  SoftMax with Threshold, OpenMax, and CROSR.

a) Unknown class detection: The performance for unknown class detection is shown in Table VIII. CRML consistently and significantly outperforms the baselines in terms of F\textsubscript{1} and Acc. across all datasets. Experimental results indicate that CRML effectively enhances the separation between known and unknown classes, effectively 
mitigating open-set risks in real-world scenarios.

b) Closed set accuracy: The ultimate goal of OSADA is not only to detect unknowns but also to maintain recognition accuracy for in-class samples. The closed set accuracy on known classes (ID Acc.) is shown in Table VIII. On the clean set, CRML's performance is slightly below that of the best baseline. However, in compressed environments, the attribution capability for known classes is enhanced, demonstrating good adaptability to complex environments.

\begin{figure}
\centering
\includegraphics[width=3in]{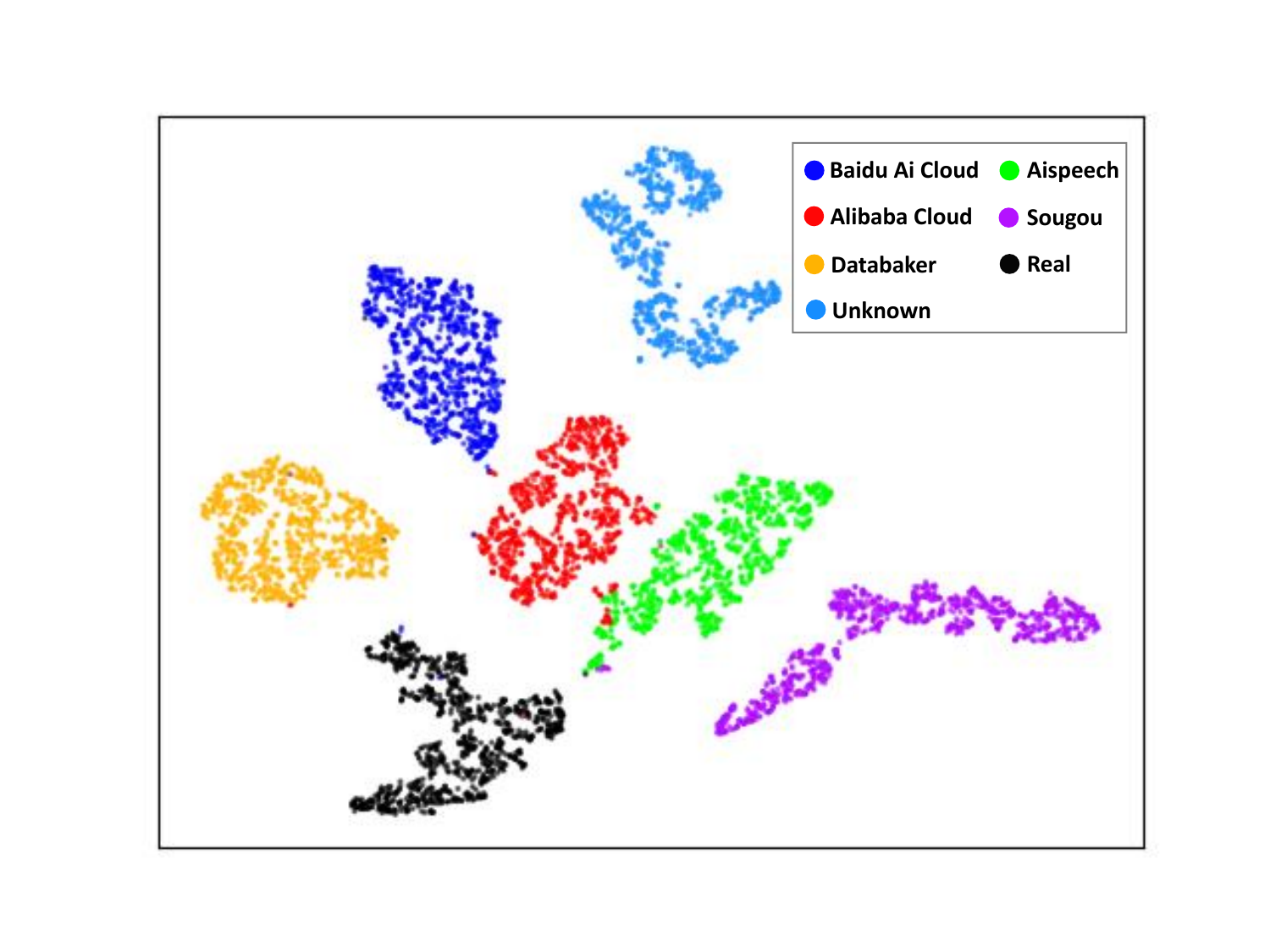}
\caption{A t-SNE visual of system fingerprint features of deepfake audio in the real-world. The result shows that differences among TTS systems can lead to effective attribution of distinct fingerprint features.}
\label{fig_1}
\end{figure}

\subsection{System Fingerprint Visualization}

Alternatively to our previous work of implicitly representing system fingerprints in the feature domain, we explicitly represent system fingerprints in the image domain. Fig 5 illustrates the frequency distribution of audio generated by five different TTS systems and a real audio sample. In each spectrogram, the horizontal axis represents time (seconds), the vertical axis represents frequency (hertz), and the color intensity indicates the energy of the audio signal at specific time and frequency points.

Due to the differences in TTS technologies among vendors, including variations in speaker corpora, divergent modelling approaches, and post-processing modules, distinctive system fingerprints are formed. These differences enable us to effectively audio deepfake attribution by leveraging the unique characteristics of each system's fingerprint.

To better illustrate the distinguishability of system fingerprints, we performed t-SNE \cite{ref47} visualization on system fingerprints in the real world. In Fig. 4, we visualize the distribution of fingerprint features from the five systems, the real audio and UNK. Different colors represent different sources. The visualization shows that system fingerprint features are a bit tangled, indicating the challenge of system fingerprint recognition oriented towards open scenarios in the real-world.

\begin{figure*}[!t]
  \centering
  \includegraphics[height=9.5cm,width=\textwidth]{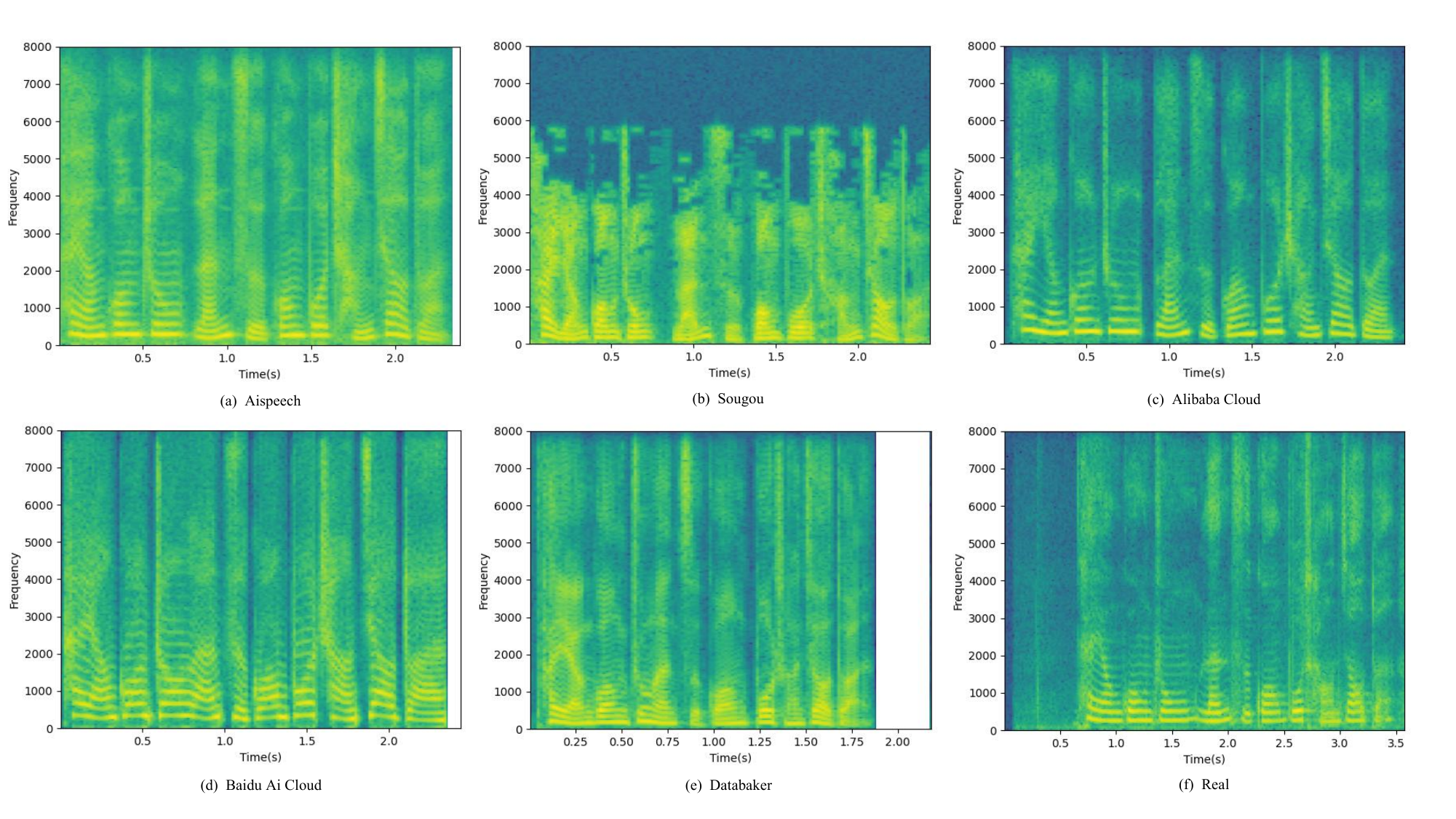}
  \caption{A spectrogram comparison of the deepfake audio from five speech synthesis systems and the real audio, the content of each audio is
  \begin{CJK*}{UTF8}{gbsn}
‘太阳光中蓝紫光波长较短.' (‘The blue-violet light in sunlight has a shorter wavelength.').
  \end{CJK*}.}
\end{figure*}

\section{Discussion}
In this section, some limitations of the present work are discussed. Both these limitations require further extended research.

The experimental results show that audio deepfake attribution is a topic that deserves further study. While encouraging, there are still some limitations in our work. We suggest future work in audio deepfake attribution along the following lines:
\begin{itemize}
\item Although ADA provides data from unknown categories, the overall number of deepfake systems used for attribution is limited. The initial dataset includes deepfake audio from synthesis systems of only seven Chinese vendors. Further research should investigate audio deepfake attribution across more vendors, languages, and voice conversion systems.
\end{itemize}

\begin{itemize}
\item Although current attribution techniques are capable of distinguishing between different audio generation tools to some extent, their effectiveness still faces significant challenges in complex scenarios involving noise interference and various languages and dialects. Future research needs to focus on developing more robust attribution algorithms.
\end{itemize}

\begin{itemize}
\item In real-world scenarios, simply rejecting unknown targets as a single “UNK” class is insufficient. There is a need to further distinguish their specific attributes. Therefore, in the future, we will focus on research related to class-incremental learning (CIL) and novel class discovery (NCD).
\end{itemize}

\section{Conclusions}
This paper provides a starting point for future research on audio deepfake attribution. We introduce a novel task of audio deepfake attribution, which aims to attribute audio generated by different deepfake technologies to their sources. This work focuses on the attribution of audio generation tools. To this end, we present the first publicly available deepfake audio dataset named ADA, which is specifically focused on the attribution of audio generation tools. We conduct an extensive investigation into the attribution of audio generation tools based on system fingerprints. The experimental results show that WavLM-SENet and wav2vec 2.0 XLS-R-SENet are relatively more suitable for audio deepfake attribution in clean and compressed environments. Moreover, we propose the class-representation multi-center learning (CRML) method for open-set audio deepfake attribution (OSADA). By optimizing the global directional variation of fingerprint representations, neural networks can more effectively learn better decision boundaries to differentiate between known and unknown classes. The experimental results demonstrate that CRML effectively addresses the attribution of unknown audio generation technologies, thereby effectively reducing the open-set risks associated with model deployment in real-world scenarios. In the future, we will focus on addressing the limitations discussed above.

\begin{IEEEbiography}[{\includegraphics[width=1in,height=1.25in,clip,keepaspectratio]{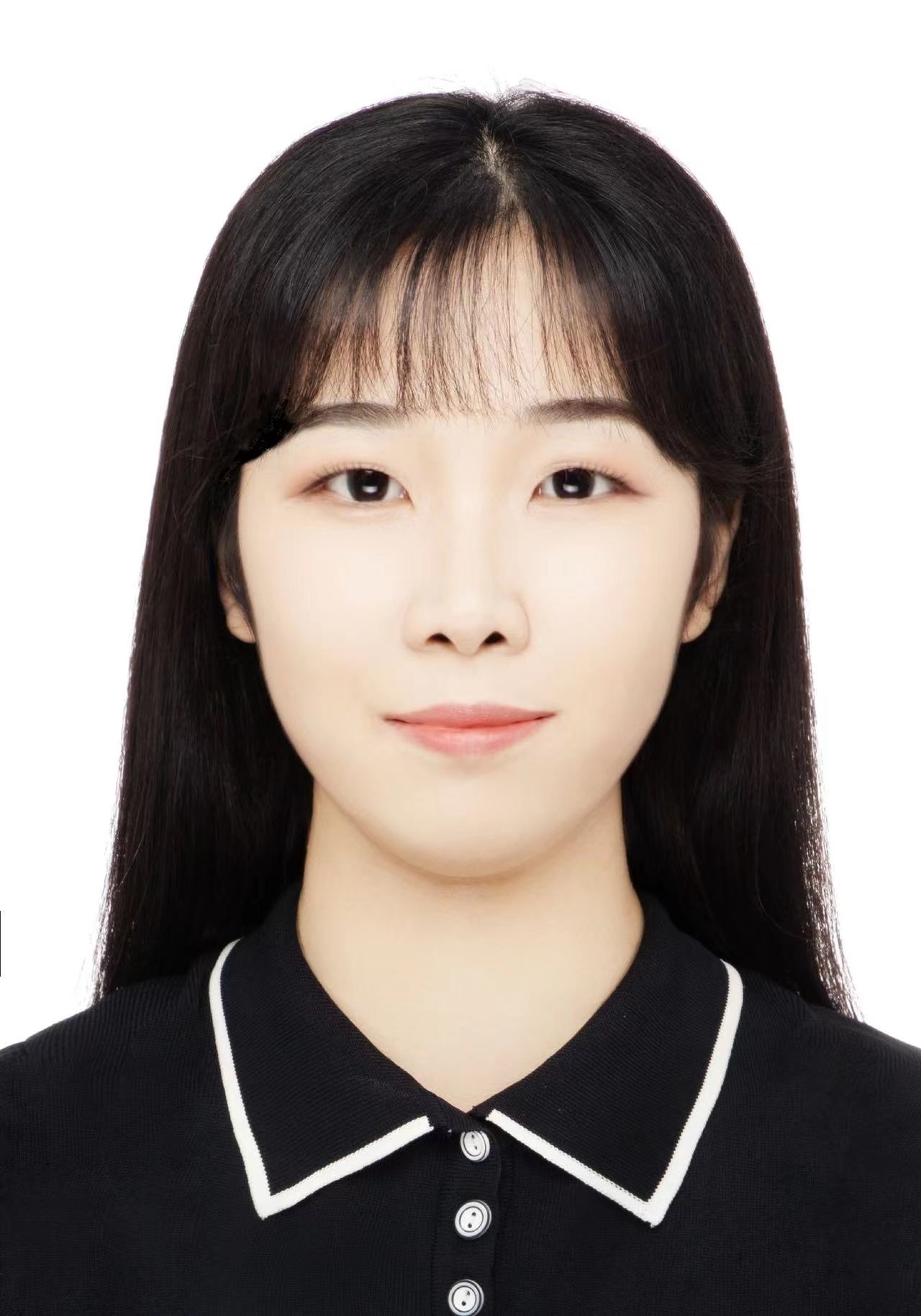}}]{Xinrui Yan}
received the B.S. degree from Northeastern University in China in 2021. She is currently pursuing her M.S. degree at the University of Chinese Academy of Sciences in Beijing, China. Her current research interest include fake audio forensics.
\end{IEEEbiography}

\begin{IEEEbiography}[{\includegraphics[width=1in,height=1.25in,clip,keepaspectratio]{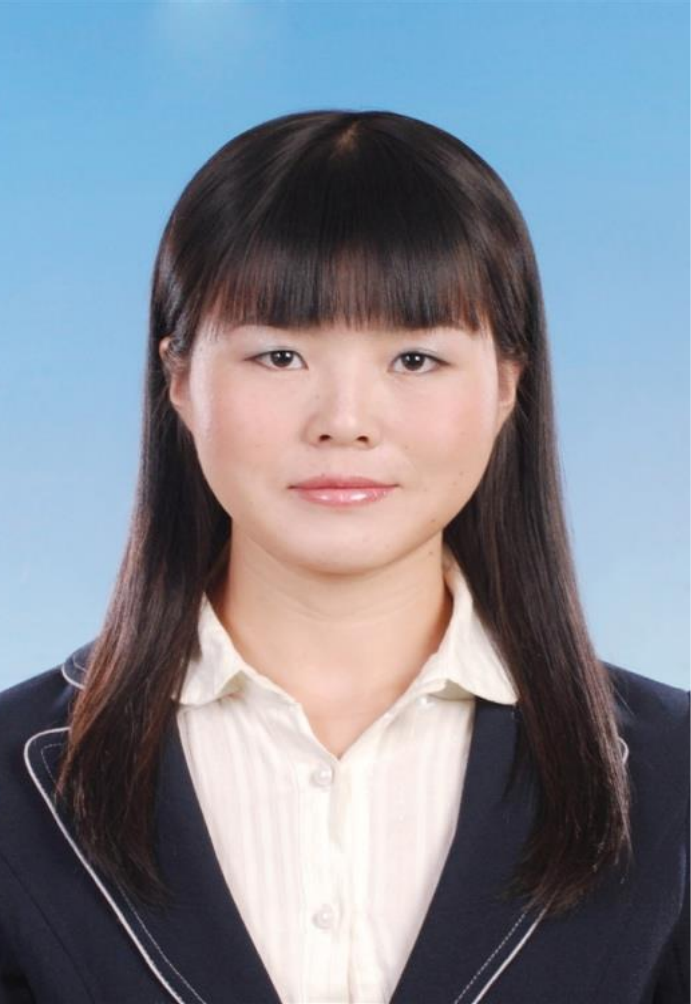}}]{Jiangyan Yi}
 received the Ph.D. degree from the University of Chinese Academy of Sciences, Beijing, China, in 2018, and the M.A. degree from the Graduate School of Chinese Academy of Social Sciences, Beijing, China, in 2010. She was a Senior R\&D Engineer with Alibaba Group during 2011 to 2014. She is currently an Associate Professor with the National Laboratory of Pattern Recognition, Institute of Automation, Chinese Academy of Sciences, Beijing, China. Her current research interests include speech signal processing, speech recognition and synthesis, fake audio detection, audio forensics and transfer learning.
\end{IEEEbiography}

\begin{IEEEbiography}[{\includegraphics[width=1in,height=1.25in,clip,keepaspectratio]{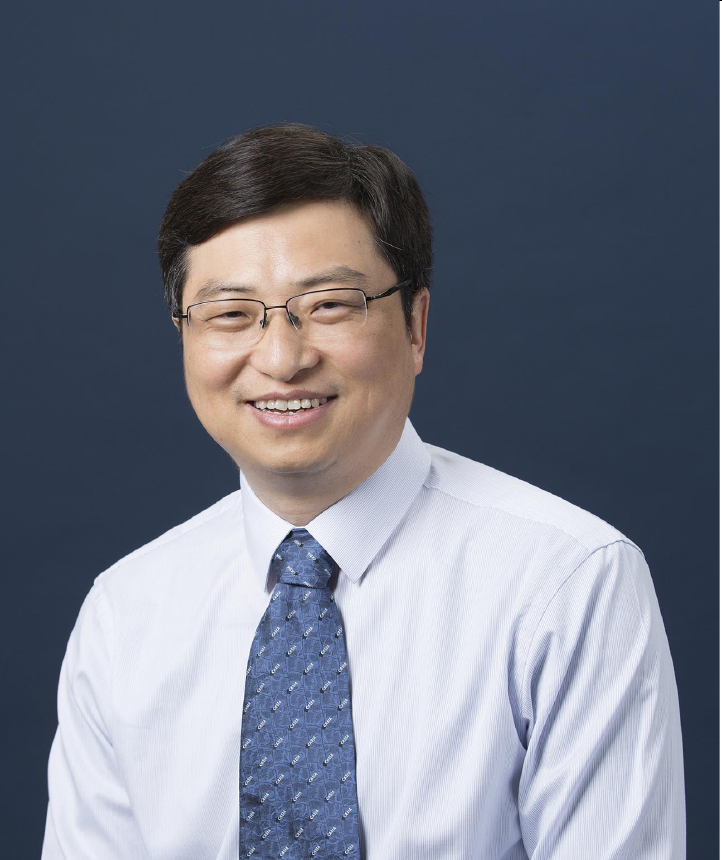}}]{Jianhua Tao}
 received his Ph.D. degree from Tsinghua University, Beijing, China, in 2001, and the M.S. degree from Nanjing University, Nanjing, China, in 1996. He is currently a Professor with Department of Automation, Tsinghua University, Beijing, China. He has authored or coauthored more than eighty papers on major journals and proceedings including the IEEE TRANSACTIONS ON AUDIO, SPEECH, AND LANGUAGE PROCESSING. His current research interests include speech signal processing, speech recognition and synthesis, human computer interaction, multimedia information processing, and pattern recognition.
\end{IEEEbiography}



\vfill

\end{CJK}
\end{document}